\documentclass[twocolumn,prb,aps,citeautoscript,amsmath,amssymb,superscriptaddress,longbibliography]{revtex4-2}

\usepackage{graphicx}
\usepackage{dcolumn}
\usepackage{bm}
\usepackage{marvosym}
\usepackage{txfonts}
\usepackage{xcolor}
\usepackage{color}
\usepackage{ulem}
\usepackage{makecell}
\usepackage{multirow}
\usepackage[colorlinks=true,urlcolor=blue,anchorcolor=blue,linkcolor=blue,citecolor=blue,breaklinks=true]{hyperref}

\begin{document}

\normalem

\newcommand{\E}{\mathrm{e}}
\newcommand{\I}{\mathrm{i}}
\newcommand{\sro}{Sr$_2$RuO$_4$}
\newcommand{\upt}{UPt$_3$}
\newcommand{\cecoin}{CeCoIn$_5$}
\newcommand{\ybco}[1]{YBa$_2$Cu$_3$O$_{#1}$}

\title{Probing time-reversal symmetry breaking at microwave frequencies}

\author{T.~Chouinard}
\author{D.~M.~Broun}
\affiliation{Department of Physics, Simon Fraser University, Burnaby, BC, V5A~1S6, Canada}

\date{\today}

\begin{abstract}
Motivated by experiments carried out in the near infrared using zero-loop-area Sagnac interferometers, we explore electromagnetic signatures of time-reversal symmetry breaking (TRSB) at microwave frequencies, using as a prototypical example a semiclassical conductor in a magnetic field.  TRSB is generically accompanied by a skew-symmetric term in the electrodynamic response tensors (permittivity, conductivity, surface impedance), imparting a nonreciprocal phase shift  to left- and right-circularly polarized electromagnetic waves  reflected from the surface of such a material.  We show that TRSB manifests as a difference in the surface reactance experienced by circularly polarized waves, and can be detected using a doubly degenerate resonator mode, such as the TE$_{111}$ mode of a cylindrical cavity.  In addition to the frequency splitting induced by TRSB we show that, when interrogated by circularly polarized microwaves, the forward and reverse transmission responses of such a resonator break reciprocity, providing a crucial signature that distinguishes true Faraday effects (i.e., circular birefringence) from non-TRSB effects such as linear birefringence.  In the limit that the sample is larger than the spot size (i.e., larger than the diameter of the microwave cavity) we show that the TRSB resonator has sensitivity to polar Kerr angle comparable to that of the zero-loop-area Sagnac, and should provide complementary insights into unconventional superconductors such as UPt$_3$ and \sro\ that have been observed to spontaneously break time-reversal symmetry.
\end{abstract}

\maketitle

\section{Introduction}

A powerful means of classifying  states of matter is by the symmetries they spontaneously break.  For example, translational and/or rotational symmetries are broken at structural phase transitions, and gauge symmetry is broken in all superconductors. As well as gauge invariance, unconventional superconductors break additional symmetries, such as the point-group symmetry that is broken in the $d$-wave pairing state of the cuprate high temperature superconductors \cite{Annett:1996wf}. A particularly subtle example, arising in superconductors such as \sro\ \cite{Luke:1998bo,Kidwingira:2006p807,Xia:2006p2} and \upt\ \cite{Luke:1993fc,Sauls:1994p138,Joynt2002,Strand:2009p137,Strand:2010p42,Schemm:2014fv}, is time-reversal-symmetry breaking (TRSB). In the case of \sro\ \cite{Mackenzie:2017do} this is believed to be associated with the existence of a pairing state such as chiral \mbox{$p$-wave} ($p_x \pm \I p_y$); chiral $d$-wave ($d_{xz} \pm \I d_{yz}$); or exotic nonchiral states such as  $d_{x^2 -y^2} \pm \I g_{xy(x^2 - y^2)}$ \cite{Kivelson:2020}, $d_{x^2 - y^2} \pm \I s^\prime$ \cite{Romer:2020} or $d_{xy} \pm \I s^\prime$ \cite{Romer:2021}, where $s'$ denotes the $A_{1g}$ symmetry $\cos(k_x ) + \cos(k_y)$ extended $s$-wave state. Although states in the latter class are not chiral, they exhibit local currents in the presence of impurities or step edges. In the case of \upt\ many experiments are well described by the $E_{2u}$ odd-parity triplet state, whose orbital part takes the chiral $f$-wave form,  $(k_x^2 - k_y^2)k_z \pm 2 \I k_x k_y k_z $ \cite{Sauls:1994p138,Joynt2002}.  It is important to note that in these TRSB combinations of order parameters, the individual gap components add in quadrature,  erasing some or all of the gap nodes that are a defining feature of unconventional pairing states such as the $d_{x^2 - y^2}$ state in the cuprates.  The $p_x \pm \I p_y$ state, for instance, is fully gapped in two dimensional systems.  This means that thermodynamic signatures of low-lying quasiparticle excitations cannot be expected to provide a definitive means of identifying the pairing state.  Instead, states that break time-reversal symmetry can be detected via the presence of spontaneous magnetic fields, and by the appearance of a spontaneous polar Kerr effect --- a difference in the interaction of the material with left-circularly polarized (LCP) and right-circularly polarized (RCP) electromagnetic fields \cite{Kapitulnik:2009p41}.  It should also be pointed out that the interpretation of experiments reporting TRSB remains controversial, and an active topic of investigation, with the possibility that in some cases TRSB may be arising from the interplay between superconductivity and disorder \cite{Andersen:2024}.

Some of the key work in this area has been carried out at near-infrared frequencies by the Kapitulnik group \cite{Xia:2006p2,Xia:2008p1,Kapitulnik:2009p41,Schemm:2014fv} with a modified, zero-loop-area Sagnac interferometer \cite{Xia:2006p36} that detects TRSB via the nonreciprocal phase shift, $\phi_\mathrm{nr}$, between LCP and RCP light reflected from the sample.  Their experiments resolve $\phi_\mathrm{nr}$ with a precision of tens of nanoradians, using state-of-the-art optical techniques to cancel out almost all common-mode phase shifts in the rest of the system. In this paper we explore electromagnetic signatures of TRSB in the microwave frequency range.  The motivation is that spectroscopic signatures of superconductivity --- i.e., changes in the optical conductivity spectrum $\sigma(\omega)$ due to the onset of superconductivity --- are generally strongest at photon energies $\hbar \omega$ comparable to the superconducting energy gap $\Delta$.  Since a thermal energy of 1~K corresponds to a frequency of approximately 20~GHz, the microwave regime should be particularly well suited to studying low $T_c$ materials such as UPt$_3$ and \sro, where the gap scale is tens of GHz.  

\section{Electromagnetic signatures of TRSB}

When using electromagnetic techniques to detect TRSB, it is important to distinguish between reciprocal effects --- e.g., the Faraday rotation observed when a linearly polarized EM wave is incident on a linearly birefringent material --- and nonreciprocal effects --- e.g., true Faraday effects due to circular birefringence, in which the sense of rotation of the polarization changes sign when the direction of the electromagnetic wave is reversed.  Particularly important here is the polar Kerr angle, $\theta_K = \phi_\mathrm{nr}/2$, which characterizes the nonreciprocal phase shift when a wave reflects from a sample that has a component of magnetization perpendicular to the sample surface, either spontaneous or induced. As shown by Argyres \cite{Argyres55}, the Kerr angle takes the form 
\begin{equation}
    \theta_K = - \mathrm{Im}\left\{\frac{\tilde n_+ - \tilde n_-}{\tilde n_+ \tilde n_- -1} \right\}\;,
    \label{eq:Argyres}
\end{equation}
where $\tilde n_+$ and $\tilde n_-$ are the complex indices of refraction for RCP and LCP electromagnetic waves.  The observation of a nonzero $\theta_K$ implies $\tilde n_+ \ne \tilde n_-$, and is an unambiguous sign that reciprocity and time-reversal symmetry have been broken in the material of interest.

\begin{figure*}[t]
\includegraphics[height = 0.35 \textwidth]{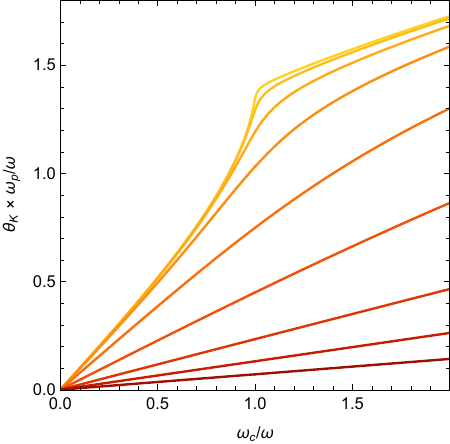} 
\qquad
\includegraphics[height = 0.35 \textwidth]{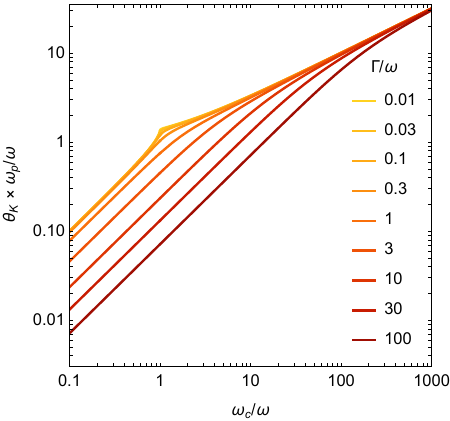}
\caption{Field dependence of the polar Kerr angle in a semi-classical conductor, for different values of the transport relaxation rate $\Gamma_\mathrm{tr}$, on linear and logarithmic axes. Applied field is parameterized by the cyclotron frequency, $\omega_c = e B/m^\ast$, in units of the measurement frequency, $\omega$.  Material-specific properties enter via the plasma frequency $\omega_p$, and are scaled out of the Kerr angle via the combination $\theta_K \times \omega_p/\omega$. }
\label{fig:Kerr_angle_semiclassical}
\end{figure*}

 In materials that display a nonzero polar Kerr effect, the permittivity and conductivity tensors develop skew-symmetric off-diagonal terms:
\begin{equation}
\epsilon = \left(\begin{array}{ccc}\tilde \epsilon_{xx} & \tilde \epsilon_{xy} & 0 \\-\tilde \epsilon_{xy} & \tilde \epsilon_{yy} & 0 \\0 & 0 & \tilde \epsilon_{zz}\end{array}\right)\;,\quad
\sigma = \left(\begin{array}{ccc}\tilde \sigma_{xx} & \tilde \sigma_{xy} & 0 \\-\tilde \sigma_{xy} & \tilde \sigma_{yy} & 0 \\0 & 0 & \tilde \sigma_{zz}\end{array}\right)\;.
\end{equation}
Without loss of generality, we can choose the  normal to the surface to be the $\hat z$ direction. For EM waves propagating into the surface, the complex polarization eigenvectors of the permittivity and conductivity tensors are $\hat e_\pm = \tfrac{1}{\sqrt{2}}(\hat x \pm \I \hat y)$.  The helicity of these states depends on whether the waves propagate in the positive or negative $\hat z$ direction, but the associated indices of refraction, $\tilde n_\pm$, (and permittivity, complex conductivity, surface impedance, etc.) remain constants of the motion, even on reflection.

In general, $\tilde n = \sqrt{\epsilon_r}$, where $\epsilon_r$ is the complex relative permittivity.  For microwave work on conducting materials it is more common to work with the complex conductivity and surface impedance, and to ignore displacement currents, which are generally negligible at GHz frequencies.  As a result, we have
\begin{equation}
    \epsilon_r \approx \frac{\sigma}{\I \omega \epsilon_0} = \frac{1}{\I \omega \epsilon_0}\frac{\I \omega \mu_0}{Z_s^2} =  \frac{\mu_0}{\epsilon_0}\frac{1}{Z_s^2} = \frac{Z_0^2}{Z_s^2}\;,
\end{equation}
where we have assumed the local-electrodynamic form for the surface impedance, and written the final expression in terms of the impedance of free space, $Z_0 = \sqrt{\mu_0/\epsilon_0} \approx 377~\Omega$.  From this we see that $\tilde n = Z_0/Z_s$. Substituting into the Argyres formula, Eq.~\ref{eq:Argyres}, we obtain
\begin{equation}
    \theta_K = \mathrm{Im}\left\{\frac{Z_s^+ - Z_s^-}{Z_0 - Z_s^+ Z_s^-/Z_0} \right\} \approx \mathrm{Im}\left\{\frac{Z_s^+ \!-\! Z_s^-}{Z_0} \right\} = \frac{X_s^+ \!-\! X_s^-}{Z_0}.
\end{equation}
Here we have dropped the second-order term $Z_s^+ Z_s^-/Z_0^2$, as it is negligible in metals, for which microwave surface impedance is typically in the m$\Omega$ range, and does not break reciprocity. The end result is that the microwave Kerr angle is the difference in surface reactance, $X_s = \mathrm{Im}\{Z_s\}$, between $\hat e_+$ and $\hat e_-$ polarizations, measured in units of $Z_0$.  A more microwave-centric derivation of the same result can be obtained from the complex reflection coefficients, $r_\pm$, of waves incident on surfaces of impedance $Z_s^\pm$ from a medium of impedance $Z_0$:
\begin{equation}
    r_\pm = \frac{Z_s^\pm - Z_0}{Z_s^\pm + Z_0}\;.
\end{equation}
The difference in reflected phase between $\hat e_+$ and $\hat e_-$ polarizations is
\begin{equation}
    \phi_\mathrm{nr} = 2 \theta_K = \mathrm{arg}\left(\frac{r_-}{r_+}\right) \approx \mathrm{arg}\left(1 + 2\frac{Z_s^+ - Z_s^-}{Z_0}\right)\;,
\end{equation}
where small, second-order terms have again been dropped.  In the regime in which $\theta_K \ll 1$, the same result is obtained:
\begin{equation}
    \theta_K \approx \mathrm{Im}\left\{\frac{Z_s^+ - Z_s^-}{Z_0} \right\} = \frac{X_s^+ - X_s^-}{Z_0}\;.
\end{equation}

\section{Semiclassical conductor in a B field}

An important example is provided by a semiclassical conductor in a magnetic field $B$, where a Kerr angle arises from the orbital motion of the charge carriers in response to the Lorentz force.  Focusing on the microwave-frequency dynamics perpendicular to the magnetic field, the transverse part of the resistivity tensor can be written
\begin{equation}
     \rho = \frac{1}{\epsilon_0 \omega_p^2} \left(\begin{array}{cc}\I \omega + \Gamma_\mathrm{tr} & - \omega_c \\+ \omega_c & \I \omega + \Gamma_\mathrm{tr}\end{array}\right)\;,
\end{equation}
where $\omega_p$ is the plasma frequency of the charge carriers and $\Gamma_\mathrm{tr}$ their transport relaxation rate. The applied field induces off-diagonal terms proportional to the cyclotron frequency, $\omega_c = e B/m^\ast$, and it is these terms that break time-reversal symmetry.  As expected, the eigenvectors of the resistivity tensor are the circularly polarized modes $\hat e_\pm$, with resistivity eigenvalues 
\begin{equation}
    \rho_\pm = \frac{1}{\epsilon_0 \omega_p^2} \Big(\I(\omega \pm \omega_c) + \Gamma_\mathrm{tr} \Big)\;.
\end{equation}
In the local electrodynamic limit,
\begin{align}
    \begin{split}
    Z_s^\pm & = \sqrt{\I \omega \mu_0 \rho_\pm}\\
    & = \I \frac{\omega}{\omega_p} Z_0 \sqrt{1 \pm \frac{\omega_c}{\omega} - \I \frac{\Gamma_\mathrm{tr}}{\omega}}\;,
    \end{split}
\end{align}
and the Kerr angle is
\begin{equation}
    \theta_K = \frac{\omega}{\omega_p}\mathrm{Re}\left\{\sqrt{1 + \frac{\omega_c}{\omega} - \I \frac{\Gamma_\mathrm{tr}}{\omega}} - \sqrt{1 - \frac{\omega_c}{\omega} - \I \frac{\Gamma_\mathrm{tr}}{\omega}} \right\}\;.
    \label{eq:Kerr_angle_semiclassical}
\end{equation}
In weak fields ($\omega_c \ll \omega$) the Kerr angle takes the form
\begin{equation}
    \theta_K \approx \frac{\omega_c}{\omega_p} \frac{1}{\sqrt{2}}\left(\frac{1 + \sqrt{1 + \Gamma_\mathrm{tr}^2/\omega^2}}{1 + \Gamma_\mathrm{tr}^2/\omega^2} \right)^{1/2}\;,
\end{equation}
with the following high and low frequency limits:
\begin{equation}
    \theta_K \approx \begin{cases}
\dfrac{\omega_c}{\omega_p}, & \omega \gg \Gamma_\mathrm{tr}, \mbox{ collisionless}\\
\dfrac{\omega_c}{\omega_p} \sqrt{\dfrac{\omega}{\Gamma_\mathrm{tr}}}, & \omega \ll \Gamma_\mathrm{tr}, \mbox{ relaxation-dominated}
\end{cases}
\end{equation}
In the collisionless regime, the Kerr angle is linear in $B$ field, inversely proportional to the plasma frequency, and independent of relaxation rate. In the relaxation-dominated regime, the leading field dependence is still linear, but decreases in scale with increasing relaxation rate.
The full field dependence of Eq.~{\ref{eq:Kerr_angle_semiclassical}} is plotted in Fig.~\ref{fig:Kerr_angle_semiclassical} on linear and logarithmic axes. In the collisionless regime, $\Gamma_\mathrm{tr}/\omega \ll 1$, the cyclotron resonance condition, $\omega_c = \omega$, appears as a kink.  In fact, in situations where the cyclotron-induced curvature is visible in the field dependence, fitting to the measured $\theta_K(B)$ should allow a simultaneous determination of the parameters $\omega_p$, $\omega_c$ and $\Gamma_\mathrm{tr}$, providing a contactless method for characterizing \mbox{magnetotransport}, and therefore giving access to important derived quantities such as carrier concentration, $n$, and Hall angle, $\theta_H = \omega_c/\Gamma_\mathrm{tr}$.

\section{Probing TRSB at Microwave frequencies}

\subsection{Circularly polarized resonator modes}
\label{sec:circ_pol_microwave_modes}

 A direct translation of the zero-loop-area Sagnac interferometer to microwave frequencies is impractical for a number of reasons, including the much longer microwave wavelength, which makes it difficult to avoid coherent interference from spurious reflections.  Instead, we can probe Kerr angle using a different type of interferometer --- the familiar microwave cavity resonator. Not only is such a system able to detect the polar Kerr effect with sensitivity similar to that of the Sagnac, but when configured in the way we propose, provides an unambiguous signature of reciprocity breaking, which is essential to properly identifying TRSB.

In order to probe the response of a sample to circularly polarized microwaves, a doubly degenerate cavity mode is required.  The TE$_{111}$ mode of a cylindrical cavity is a particularly good choice, being well separated from other cavity modes, and having mode patterns reminiscent of linearly-polarized light. In the $z$ basis, as shown in Fig.~\ref{fig:TE111_modes}, these are a vertical polarization, $|0\rangle =  \left(\begin{smallmatrix}1\\0\end{smallmatrix}\right)$, and a horizontal polarization, $|1\rangle = \left(\begin{smallmatrix}0\\1\end{smallmatrix} \right)$.  When a sample that breaks time-reversal symmetry is placed inside such a resonator, it lifts the degeneracy of the two resonator modes and forms new eigenstates that are circularly polarized superpositions of the original modes:
\begin{equation}
    |\pm\rangle = \tfrac{1}{\sqrt{2}}\big(|0\rangle \pm \I |1\rangle \big)\;.
\end{equation}

\subsection{Resonator perturbation}

According to the theory of cavity perturbation \cite{Ormeno:1997p342,Huttema:2006p344}, the shift in resonant frequency of each circularly polarized mode contains a contribution from the corresponding surface reactance of the sample,   
\begin{equation}
    \Delta \omega_\pm = -\Gamma \, X_s^\pm\;.
\end{equation}
Here, the geometric factor (resonator constant) $\Gamma$ is determined by the size and shape of the sample, and the microwave magnetic field profile $H(\mathbf{r})$ in the mode of interest: 
\begin{equation}
    \Gamma = \frac{1}{2 \mu_0} \left.\int_S H^2 dS\middle/\right.\int_V H^2 dV\;,
    \label{eq:gamma1}
\end{equation}
where the integrals denote integration over the sample surface and cavity volume, respectively.  It is common to place the sample at a local maximum of $|H(\mathbf{r})|$, which for the TE$_{111}$ mode would be at the center of one of the end walls.  If the sample is small enough, then $H(\mathbf{r})$ will be approximately constant over the surface of the sample, so that the surface integral in Eq.~\ref{eq:gamma1} is well approximated by $H_\mathrm{sample}^2 \times A_\mathrm{sample}$, where $A_\mathrm{sample}$ is the total area of the sample, including top and bottom faces.  It is then useful to define an effective resonator volume for the microwave mode of interest:
\begin{figure}[t]
\includegraphics[width = 0.95 \columnwidth]{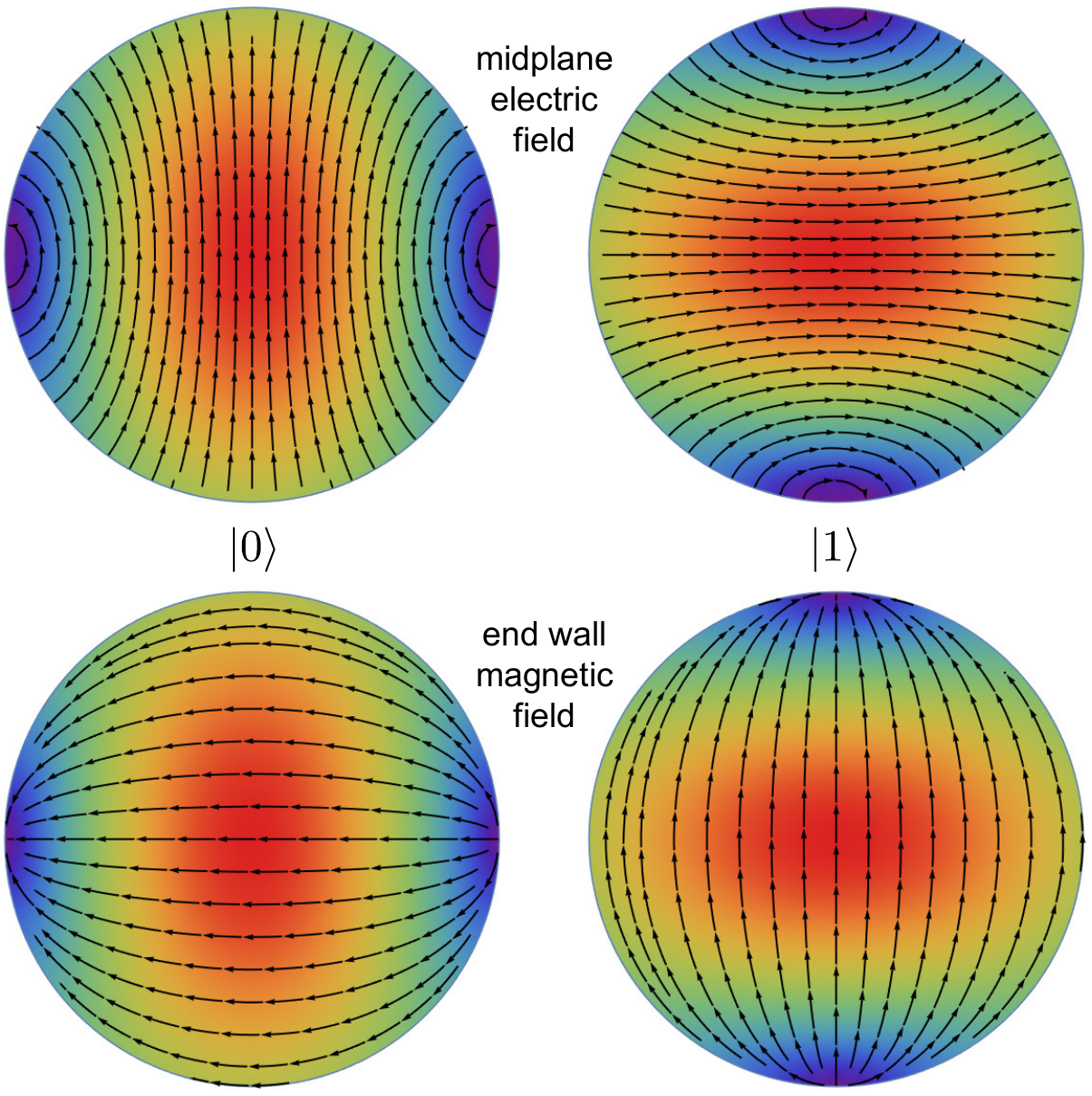}
\caption{Electric and magnetic fields of the doubly degenerate TE$_{111}$ mode of a cylindrical cavity, which serve as the linearly polarized basis states $|0\rangle$ (left column) and $|1\rangle$ (right column) in the TRSB resonator.  The top row plots the electric field at the midplane of the resonator, in each of the modes, with electric field strength highest at the center of the cavity.  The bottom row plots the magnetic field at the end walls of the resonator: magnetic field strength is highest at the center of the end walls.}
\label{fig:TE111_modes}
\end{figure}
\begin{figure*}[t]
\includegraphics[width = 0.95 \textwidth]{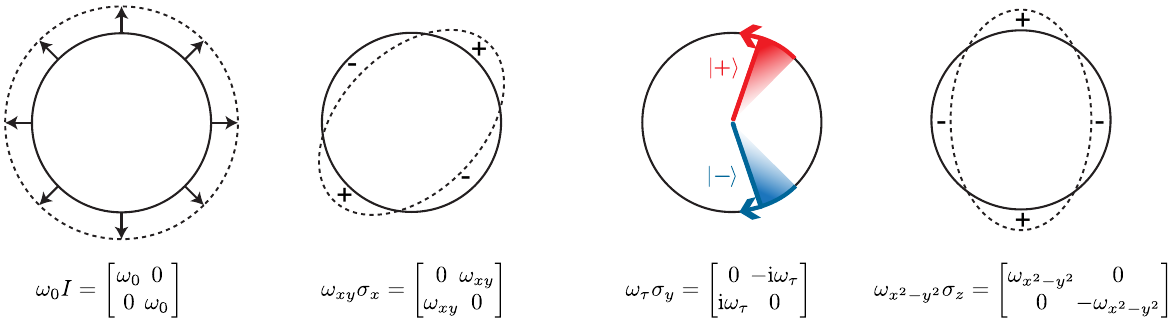}
\caption{The four types of perturbation to the TE$_{111}$ resonator, each labeled by its corresponding matrix perturbation in the $z$ basis. $\omega_o I$: trivial, common-mode perturbations, such as an isotropic resonator stretch, which affect both polarizations equally; $\omega_{xy}\sigma_x$: $xy$-type quadrupolar distortions, such as a deformation of the resonator cross section aligned with the diagonal axes; $\omega_\tau\sigma_y$: time-reversal-symmetry-breaking perturbations, which have opposite effect on left- and right-precessing modes; and $\omega_{x^2 - y^2}\sigma_z$: $x^2 - y^2$ quadrupolar perturbations such as a deformation of the resonator cross section aligned with the horizontal and vertical axes.}
\label{fig:perturbations}
\end{figure*}
\begin{equation}
    V_\mathrm{eff} = \frac{1}{H_\mathrm{sample}^2} \int_V H^2 dV\;.
\end{equation}
As shown in Appendix~\ref{sec:appendix}, for a sample positioned at the center of the end wall of a TE$_{111}$ cavity, and for typical aspect ratio, $V_\mathrm{eff} \approx V_\mathrm{cavity}$, where $V_\mathrm{cavity} = \pi a^2 d$ is the physical volume of the cylindrical resonator.  In terms of this, the resonator constant can be written
\begin{equation}
    \Gamma = \frac{1}{2 \mu_0} \frac{A_\mathrm{sample}}{V_\mathrm{eff}}\;.
    \label{eq:gamma2}
\end{equation}
Applying this to the $|\pm\rangle$ modes of the cavity, the relative frequency shift can be related to the Kerr angle:
\begin{equation}
    \Delta\omega_+ - \Delta\omega_- = \Gamma \left(X_s^+ - X_s^-\right) = \Gamma\, Z_0 \theta_K\;.
    \label{eq:splitting}
\end{equation}
Or, rearranging,
\begin{align}
\begin{split}
    \theta_K & = \frac{\Delta\omega_+ - \Delta\omega_-}{\Gamma Z_0}\\
    & = \frac{\mu_0}{Z_0} \frac{2 V_\mathrm{eff}}{A_\mathrm{sample}}\left(\Delta\omega_+ - \Delta\omega_-\right)\\
    & = \frac{2 V_\mathrm{eff}}{A_\mathrm{sample}}\frac{1}{c} \left(\Delta\omega_+ - \Delta\omega_-\right)\\
    & = \frac{2 V_\mathrm{eff}}{A_\mathrm{sample}}2 \pi \left[\Delta\left(\frac{1}{\lambda_+}\right) - \Delta\left(\frac{1}{\lambda_+}\right)\right]\\
    & = \frac{2 V_\mathrm{eff}}{A_\mathrm{sample} \lambda}2 \pi \left(\frac{\Delta \lambda_+}{\lambda} - \frac{\Delta \lambda_-}{\lambda}\right)\;.
    \end{split}
\end{align}
where $c$ is the speed of light in vacuum, $\lambda_\pm$ are the free space wavelengths associated with each mode, and $\lambda$ is the average of $\lambda_\pm$.  In the limit where the sample forms one complete end wall of the resonator we would arrive at something very similar to the last line just on the basis of the nonreciprocal phase shift.  In this limit, the cavity mode consists of microwaves reflecting back and forth between end walls separated by a distance $\lambda/2$ (i.e., a Fabry--P\'erot cavity), and $2 V_\mathrm{eff}/A_\mathrm{sample} \lambda \approx \frac{1}{2}$.  The phase shift of each mode is $2 \pi$ times the fractional change in wavelength (e.g., $\Delta \phi_+ = 2 \pi \Delta \lambda_+/\lambda$), and can therefore be directly related back to the nonreciprocal phase shift, $\phi_\mathrm{nr} = \Delta \phi_+ - \Delta \phi_- = 2 \theta_K$.

While all the information about Kerr angle is contained in the relative frequency shift $\Delta\omega_+ - \Delta\omega_-$, in practice it is not possible to simply measure the frequency splitting and assign it to TRSB. This is because distortions of the resonator shape (i.e., deviations from cylindrical symmetry) cause spurious frequency shifts that are unrelated to TRSB.  In addition, dissipation in the resonator imparts a finite resonant bandwidth to the cavity modes,  making it impossible to separately resolve near-degenerate modes when their splitting is much smaller than the resonant bandwidth.  Finally, the observation of frequency splitting on its own does not indicate whether reciprocity is broken, a test that should always accompany a measurement of TRSB. The solution to these problems requires a more subtle approach, based on a formal treatment of the two-level system formed by the pair of near-degenerate cavity modes, which we carry out in the next section.

\subsection{Perturbations of the two-mode system}

In the frequency range of interest, the Hilbert space of the near-degenerate TE$_{111}$ resonator is spanned by the two orthogonal linear polarizations shown in Fig.~\ref{fig:TE111_modes} (the $z$ basis). Let the photon creation operators for these modes be $a^\dagger$ and $b^\dagger$, respectively. The most general Hamiltonian for the two-level system can be written in terms of spinors as 
\begin{equation}
    (a^\dagger, b^\dagger) \cdot H \cdot \left(\!\!\begin{array}{c}a \\b\end{array}\!\!\right)\;,
\end{equation}
where the system matrix is
\begin{equation}
    H = H_0 + H_1 = \omega_0 I + \omega_{xy} \sigma_x + \omega_\tau \sigma_y + \omega_{x^2 - y^2} \sigma_z\;.
\end{equation}
 Here $I$ is the $2 \times 2$ identity matrix and the $\sigma_i$ are the Pauli matrices.  From this, we see that the set of all nontrivial perturbations to the degenerate two-mode system can be written compactly as $H_1 = \vec \Omega \cdot \vec \sigma$, where $\vec \Omega = (\omega_{xy},\omega_\tau,\omega_{x^2 - y^2})$ and $\vec \sigma = (\sigma_x,\sigma_y,\sigma_z)$.  The system is therefore analogous to the well-known problem of a spin-$\tfrac{1}{2}$ particle in a magnetic field.  
 
Two of the perturbations correspond to shape distortions of the resonator:
\begin{align}
    \omega_{xy} \sigma_x & = \left[\begin{array}{cc}0 & \omega_{xy} \\\omega_{xy} & 0\end{array}\right]\;,\\
    \omega_{x^2-y^2} \sigma_z & = \left[\begin{array}{cc}\omega_{x^2-y^2} & 0 \\0 & -\omega_{x^2-y^2}\end{array}\right]\;.
\end{align}
They are labelled $xy$ and $x^2 - y^2$ because in the two-mode system, the effect of any deviation from cylindrical symmetry can be represented as the combined effect of two independent quadrupolar distortions of the resonator boundary, as illustrated in Fig~\ref{fig:perturbations}.  This includes perturbations caused by an arbitrary-shaped sample, and any deviation of sample position away from the symmetry axis of the cylindrical resonator.

The quantity of interest in the Kerr effect measurement is the perturbation that breaks time-reversal symmetry:
\begin{equation}
    \omega_\tau \sigma_y = \left[\begin{array}{cc}0 & - \I \omega_\tau \\ \I \omega_\tau & 0\end{array}\right]\;.
\end{equation}
As discussed above, it arises from the presence of skew-symmetric off-diagonal terms in the electrodynamic tensors and, on its own, results in circularly polarized $|\pm\rangle$ eigenstates.  
In the presence of perturbations of all three types, the resonator spectrum has the Dirac form:
\begin{align}
    \begin{split}
    \omega &= \omega_0 \pm | \Omega | \\
        & = \omega_0 \pm \sqrt{\smash{\omega^2_{xy} + \omega_\tau^2 + \omega_{x^2 - y^2}^2} \vphantom{|}}\;.
    \end{split}
\end{align}
\begin{figure*}[t]
\includegraphics[width = 0.95 \textwidth]{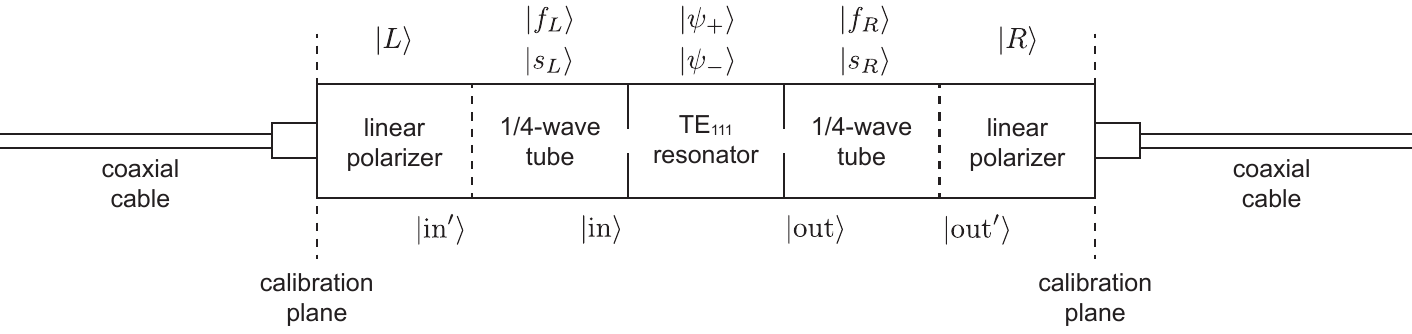}
\caption{Schematic of the TRSB resonator system showing how in the forward direction: microwaves are launched from the left-hand coaxial cable into a short section of waveguide as a linearly polarized mode; are incident on a quarter-wave tube that converts from linear to circular polarization; are coupled into the $TE_{111}$ resonator through an on-axis coupling iris; are coupled out of the resonator into a second quarter-wave tube that converts circular back into linear polarization; and are analyzed by a linear polarizer before being launched back onto a second coaxial cable.  The local eigenstates in each section are shown above the assembly.  The local input and output states, which are superpositions of the local eigenstates, are shown below the assembly.  When the system is operated in reverse, the sense of the circular polarization changes.  In the absence of TRSB, the resonator system is reciprocal: i.e., the forward and reverse transmission response is the same.}
\label{fig:resonator_schematic}
\end{figure*}
We see from the Dirac-cone part of the spectrum that in the absence of quadrupolar distortions, the frequency splitting indeed corresponds to the degree of TRSB: $\omega = \omega_0 \pm \omega_\tau$.  Combining with Eq.~\ref{eq:splitting}, we arrive at our central result for the Kerr angle:
\begin{equation}
\theta_K = \frac{2 \omega_\tau}{\Gamma\,Z_0}\;.
\end{equation}
Note that in this expression, $\omega_\tau$ appears in units of radians/second.  This will be important for the quantitative estimates of measurement sensitivity in Sec.~\ref{sec:sensistivity}.

For the reasons noted above, a measurement of the frequency splitting alone is not adequate as a  strategy for detecting TRSB, not least because it does not provide a test of reciprocity breaking. A better approach is to interrogate the resonator with left- and right-circularly polarized microwaves, in which case the presence of an $\omega_\tau \sigma_y$ term (TRSB) is revealed as a breaking of reciprocity (i.e., a difference between forward and reverse transmission through the two-port resonator).  It is this strategy that we describe below.

\subsection{Transmission response and reciprocity breaking}

A resonator system to detect TRSB at microwave frequencies is shown schematically in Fig.~\ref{fig:resonator_schematic}, and contains several different sections.  At the heart of the system is the two-mode resonator ––– in our preferred design a cylindrical cavity supporting a pair of TE$_{111}$ modes.  In order to interrogate this resonator with circularly polarized microwaves, a coaxial cable launches a linearly polarized TE$_{11}$ mode into a short section of circular waveguide.  This is connected to a quarter-wave tube, in which the fast and slow axes are oriented at 45$^\circ$ to the linearly polarized input wave.  At the far side of the quarter-wave tube, a circularly polarized TE$_{11}$ mode emerges.  This is coupled into the first port of the resonator through an on-axis coupling iris, where it excites a superposition of the resonator's two near-degenerate TE$_{111}$ eigenmodes, $|\psi_+\rangle$ and $|\psi_-\rangle$, as determined by the particular perturbations $\vec \Omega$ present in the resonator.  At the second port of the resonator, microwave energy is coupled back out into a second quarter-wave tube and converted back into linearly polarized microwaves.  The output of the quarter-wave tube is connected to a second linear polarizer, which carries out a projective sampling of the output wave before launching an output signal onto a second coaxial cable.  Implicit here is that the input port of the resonator is being driven harmonically at an angular frequency $\omega$, imparting a common time and space dependence $\E^{\I (\omega t - k_z z)}$ to all fields. Synchronous detection, at the same frequency, occurs at the output, and is carried out using a microwave vector network analyzer (VNA), with its calibration planes set to coincide with the left and right ports of the resonator assembly.  The measured response is therefore the transmission amplitude through the resonator and couplers, measured by the VNA as a complex phasor amplitude.

In each section of the resonator assembly we define the operations that take place (launching, phase-shifting, resonant coupling and projection) in terms of the local eigenbasis, as indicated at the top of Fig.~\ref{fig:resonator_schematic}.  The goal is to calculate the transmission amplitude through the distributed system, starting at the left-hand side with an input state $|\mathrm{in}^\prime \rangle$ = $|L\rangle$, and then finishing by projecting the output state $|\mathrm{out}^\prime \rangle$ onto the right-hand polarizer state $|R\rangle$ to obtain the complex transmission amplitude.  Reverse propagation is obtained by switching $L \leftrightarrow R$.  The necessary transformations in between are carried out using a sequence of projection operators.

Propagation from left to right begins by decomposing the input state into the fast and slow modes of the left-hand quarter-wave tube:
\begin{equation}
    |\mathrm{in}^\prime\rangle = |L\rangle = \big(|f_L\rangle \langle f_L | + |s_L\rangle \langle s_L | \big)|L\rangle\;.
\end{equation}
The left-hand quarter-wave tube imparts a relative phase shift $\theta_L$ to the slow mode, so that the state entering the resonator is 
\begin{equation}
    |\mathrm{in}\rangle = \big(|f_L\rangle \langle f_L | + \E^{\I \theta_L}|s_L\rangle \langle s_L | \big)|L\rangle\ \equiv Q_L(\theta_L) |L\rangle\;.
\end{equation}
We use this to define the operator representation of the left-hand quarter-wave tube,  $Q_L(\theta_L)$, with a similar operator $Q_R(\theta_R)$ on the right.  By considering arbitrary phase shifts $\theta_L$ and $\theta_R$, we can capture the effects of elliptical polarization, thereby allowing for imperfections in the quarter-wave tubes.

We provide a general description of the resonator eigenmodes, $|\psi_+\rangle$ and $|\psi_-\rangle$, in the next section, using the standard theory of two-level systems.  Each eigenmode provides an independent path for resonantly coupling through the system, with scattering parameters of the form
\begin{equation}
    \tilde S_{21}^\pm = \frac{S_0^\pm}{1 + 2 \I (f - f_0^\pm)/f_B^\pm}\;,
\end{equation}
where, for the $|\psi_\pm\rangle$ modes respectively, $f_0^\pm$ are the resonant frequencies; $f_B^\pm$  the resonant bandwidths; and $\tilde S_0^\pm$ the complex scattering parameters on resonance, built in to which are the effects of weakly coupling through the input and output irises of the resonator \footnote{In addition to the resonant coupling channels it is sometimes also necessary to consider direct coupling between input and output ports.  This can be modelled via the addition of a complex amplitude $\tilde D_{21}$.}.  The output state of the resonator can then be expressed in terms of the projection operators \mbox{$|\psi_\pm\rangle\langle \psi_\pm|$}:
\begin{equation}
    |\mathrm{out}\rangle = \left(\tilde S_{21}^+(f) |\psi_+\rangle\langle \psi_+| + \tilde S_{21}^-(f) |\psi_-\rangle\langle \psi_-|\right) |\mathrm{in}\rangle\;.
\end{equation}
The state $|\mathrm{out}\rangle$ enters the right-hand quarter-wave tube, and the relative phase shift $\theta_R$ is imparted along the slow axis:
\begin{equation}
    |\mathrm{out}^\prime\rangle = \big(|f_R\rangle \langle f_R | + \E^{\I \theta_R}|s_R\rangle \langle s_R | \big)|\mathrm{out}\rangle\;.
\end{equation}
The state $|\mathrm{out}^\prime\rangle$ is then projected onto the right-hand polarizer to obtain the forward transmission through the system, \mbox{$\tilde S_{21}(f) = \langle R | \mathrm{out}^\prime \rangle$}.

In analyzing the experiment, we are particularly interested in the complex amplitude factors that multiply the individual resonances, which we define in the forward (i.e.,  $|L\rangle \to |R\rangle$) direction, as:
\begin{equation}
        A_\pm  = \langle R | Q_R(\theta_R) |\psi_\pm \rangle \langle \psi_\pm | Q_L(\theta_L) | L\rangle\;.
   \label{eq:forward_amplitudes}
\end{equation}
In the reverse ($|R\rangle \to |L\rangle$) direction these are
\begin{equation}
        \overline A_\pm  = \langle L | Q_L(\theta_L) |\psi_\pm \rangle \langle \psi_\pm | Q_R(\theta_R) | R\rangle\;.
    \label{eq:reverse_amplitudes}
\end{equation}
The quarter-wave operators have the general form 
\begin{equation}
    Q = |f\rangle \langle f| + \E^{\I \theta} |s\rangle \langle s|\;.
\end{equation}
The presence of the phase factor associated with the slow mode means that the $Q$ operators are not hermitian, but they are symmetric: $Q = Q^\top$. Additionally, the linear polarizations, $|L\rangle$ and $|R\rangle$, can always be chosen to be real, so that taking the transpose and the hermitian adjoint have the same effect. We can therefore relate the forward and reverse transmission amplitudes using the transpose operator.  That is,
\begin{equation}
    \begin{split}
        (A_\pm)^\top & = \big( \langle R | Q_R(\theta_R) |\psi_\pm \rangle \langle \psi_\pm | Q_L(\theta_L)  | L\rangle\big)^\top\\
        & = \langle L |^\ast Q_R^\top(\theta_R) \big(|\psi_\pm \rangle \langle \psi_\pm |\big)^\top Q_L^\top(\theta_L) | R\rangle^\ast \\
        & = \langle L | Q_R(\theta_R) \big(|\psi_\pm \rangle \langle \psi_\pm |\big)^\top Q_L(\theta_L) | R\rangle\;.
    \end{split}
\end{equation}
We see that the transmission response of the resonator will be reciprocal (i.e., $A_\pm = \overline A_\pm$) if and only if the following reciprocity condition holds:
\begin{equation}
    |\psi_\pm \rangle \langle \psi_\pm | = \big(|\psi_\pm \rangle \langle \psi_\pm |\big)^\top\;.
    \label{eq:reciprocity_condition}
\end{equation}

\subsection{Resonator eigenstates and signatures of TRSB}

To explore the effect of general perturbations to the TRSB resonator, we parameterize the set of all possible perturbations in spherical coordinates, using angles $\theta$ and $\phi$ that trace out the set of points on the unit sphere (the Bloch sphere).  In terms of these, the vector of perturbations can be written
\begin{align}
    \begin{split}
    \vec \Omega &= 
    \left(\omega_{xy}, \omega_\tau, \omega_{x^2 - y^2}\right)\\
    &= |\Omega|\,\big(\sin \theta \cos \phi, \sin \theta \sin \phi, \cos \theta\big)\;.
    \end{split}
\end{align}
The general form of the perturbation Hamiltonian is then
\begin{equation}
    H_1 = \vec \Omega \cdot \vec \sigma = |\Omega|\left[\begin{array}{cc}\cos \theta & \sin \theta \E^{- \I \phi} \\\sin \theta \E^{\I \phi} & -\cos \theta\end{array}\right]\;.
\end{equation}
The eigenstates of $H_1$ are well known from the theory of the spin-$\tfrac{1}{2}$ particle in a magnetic field and can be written
\begin{equation}
    |\psi_+\rangle = \left(\renewcommand{\arraystretch}{1.3}\begin{array}{c}\cos \frac{\theta}{2} \E^{-\I\phi/2} \\\sin \frac{\theta}{2} \E^{\I \phi/2}\end{array}\right),\;|\psi_-\rangle = \left(\renewcommand{\arraystretch}{1.3}\begin{array}{c}-\sin \frac{\theta}{2} \E^{-\I\phi/2} \\\cos \frac{\theta}{2} \E^{\I \phi/2}\end{array}\right)\;.
\end{equation}
The projection operators follow from these
\begin{align}
        |\psi_+ \rangle \langle \psi_+ | &  = \left[\renewcommand{\arraystretch}{1.3}\begin{array}{cc}\cos^2 \frac{\theta}{2} & \E^{-\I \phi}\sin \frac{\theta}{2}\cos\frac{\theta}{2} \\ \E^{\I \phi}\sin \frac{\theta}{2}\cos\frac{\theta}{2} & \sin^2 \frac{\theta}{2}\end{array}\right],\\
        |\psi_- \rangle \langle \psi_- | &  = \left[\renewcommand{\arraystretch}{1.3}\begin{array}{cc}\sin^2 \frac{\theta}{2} & -\E^{-\I \phi}\sin \frac{\theta}{2}\cos\frac{\theta}{2} \\ -\E^{\I \phi}\sin \frac{\theta}{2}\cos\frac{\theta}{2} & \cos^2 \frac{\theta}{2}\end{array}\right]\;,
\end{align}
which can be combined to give
\begin{equation}
        \begin{split}
        |\psi_\pm \rangle \langle \psi_\pm | & = \tfrac{1}{2} \left(I \pm \left[\begin{array}{cc}\cos \theta & \sin\theta \E^{-\I\phi} \\ \sin\theta \E^{\I\phi} & -\cos \theta\end{array}\right] \right)\\
        & = \tfrac{1}{2} \left(I \pm \frac{1}{|\Omega|} \left[\begin{array}{cc}\omega_{x^2+y^2} & \omega_{xy} - \I \omega_\tau \\ \omega_{xy} + \I \omega_\tau & - \omega_{x^2+y^2}\end{array}\right] \right)\\
        & = \tfrac{1}{2} \left(I \pm \frac{1}{|\Omega|}\left(\omega_{xy} \sigma_x + \omega_\tau \sigma_y + \omega_{x^2 + y^2} \sigma_z  \right) \right)\;.
        \end{split}
    \label{eq:projection_operators}
\end{equation}
We can now see the connection between TRSB and the breaking of reciprocity --- the $\omega_\tau \sigma_y$ term in Eq.~\ref{eq:projection_operators} is the only antisymmetric term, and therefore the only one to violate the reciprocity condition, Eq.~\ref{eq:reciprocity_condition}.  Furthermore, if we now specialize to the limit described in the next section, where the polarizers and quarter-wave-tubes have been tuned to produce circularly polarized microwaves, we  obtain simple expressions for the resonant amplitudes. The circularly polarized $|+\rangle$ and $|-\rangle$ states are eigenstates of the $\sigma_y$ Pauli matrix, with eigenvalues $\pm 1$, and $\sigma_y$ has expectation values
\begin{equation}
    \langle \pm | \sigma_y | \pm \rangle = \pm 1\;.
\end{equation}
Similarly, the $\sigma_x$ and $\sigma_z$ Pauli matrices have zero expectation value in the $|\pm\rangle$ states:
\begin{equation}
    \langle \pm | \sigma_{x,z} | \pm \rangle = 0\;.
\end{equation}
The forward transmission amplitudes through the two-mode resonator are then
\begin{equation}
\begin{split}
    A_\pm  &= \langle R | Q_R(\theta_R) |\psi_\pm \rangle \langle \psi_\pm | Q_L(\theta_L) | L\rangle\\
    & = \langle + |\psi_\pm \rangle \langle \psi_\pm | + \rangle\\
    & = \tfrac{1}{2} \big(1 \pm \omega_\tau/|\Omega| \big)\;.
\end{split}
\end{equation}
The reverse transmission amplitudes are
\begin{equation}
\begin{split}
    \overline A_\pm  & = \langle L | Q_L(\theta_L) |\psi_\pm \rangle \langle \psi_\pm | Q_R(\theta_R) | R\rangle\\
    & = \langle - |\psi_\pm \rangle \langle \psi_\pm | - \rangle\\
    & = \tfrac{1}{2} \big(1 \mp \omega_\tau/|\Omega| \big)\;.
\end{split}
\end{equation}
That is, the magnitude of the TRSB term $\omega_\tau$, relative to the total frequency splitting $|\Omega|$, is given by the left--right asymmetry in amplitude of the two microwave modes in the forward direction, with the sense of the asymmetry reversed when the direction of the microwaves is reversed.  This breaking of reciprocity confirms that the effect is due to TRSB.

In order to quantify the magnitude of the $\omega_\tau$ term, it is useful to define a pair of metrics that capture the left--right asymmetry and the forward--reverse asymmetry that TRSB induces in the resonant amplitudes.  When the coupling to the two microwave modes is highly symmetric, a subtractive measure can be defined as
\begin{equation}
   \left(A_+ - A_-\right) - \left(\overline A_- - \overline A_+\right) = 2 \omega_\tau/|\Omega|\;.
   \label{eq:subtractive_measure}
\end{equation}
However, when there is asymmetry of coupling to the two microwave modes, which is unavoidable in practice due to imperfections, e.g., in the coupling irises, a ratiometric measure is more versatile:
\begin{equation}
    \frac{A_+}{A_-}\times \frac{\overline A_-}{\overline A_+} = \frac{\left(1 + \omega_\tau/|\Omega|\right)^2}{\left(1 - \omega_\tau/|\Omega|\right)^2}\;.
    \label{eq:ratiometric_measure}
\end{equation}
This has the advantage that any asymmetries in coupling cancel out.  In the limit of weak TRSB, $\omega_\tau \ll |\Omega|$, the ratiometric measure goes to $1 + 4 \omega_\tau/|\Omega|$.

\section{Measuring strong and weak TRSB}

In this section we illustrate the operation of the TRSB resonator system in the important asymptotic cases of strong and weak TRSB.  We start by choosing particular representations for the polarizers, and by making some simplifying assumptions about the quarter-wave tubes.

\subsection{Interrogating with circularly polarized microwaves}

Without loss of generality, we assume that the linear polarizers are oriented at 45$^\circ$ with respect to the $|0\rangle$ and $|1\rangle$ states of the $z$ basis, and at 90$^\circ$ with respect to each other: \mbox{$|L\rangle = \frac{1}{\sqrt{2}} \left(\begin{smallmatrix}1\\1\end{smallmatrix}\right)$} and \mbox{$|R\rangle = \frac{1}{\sqrt{2}} \left(\begin{smallmatrix}1\\-1\end{smallmatrix}\right)$}. We will see that the 90$^\circ$ relative alignment is important, as it allows for the fact that microwaves passing  through two quarter-wave tubes in series are rotated by a total angle of 90$^\circ$.  The fast and slow axes of the circular polarizers can then be aligned with the $z$ basis, with \mbox{$|f\rangle = |0\rangle =  \left(\begin{smallmatrix}1\\0\end{smallmatrix}\right)$} and \mbox{$|s\rangle = |1\rangle =  \left(\begin{smallmatrix}0\\1\end{smallmatrix}\right)$}, respectively. For simplicity, we assume that the quarter-wave tubes are perfectly tuned: i.e, that \mbox{$\theta_L = \theta_R = \pi/2$}.  Deviations from the quarter-wave condition cause no sudden breakdown of the measurement --- it is still possible to probe the TRSB signal by interrogating the resonators with elliptically polarized microwaves --- there is just a gradual reduction in sensitivity away from the quarter-wave condition.  The quarter-wave-tube projection operators take the form 
\begin{equation}
    Q_L = Q_R = |f\rangle \langle f| + \I |s\rangle \langle s| = \left[\renewcommand{\arraystretch}{1.0}\begin{array}{cc}1 & 0 \\0 & \I\end{array}\right]\;.
\end{equation}
In the forward direction, we see that 
\begin{align}
    Q_L |L\rangle & = \left[\renewcommand{\arraystretch}{1.0}\begin{array}{cc}1 & 0 \\0 & \I\end{array}\right] \cdot \tfrac{1}{\sqrt{2}}\left(\renewcommand{\arraystretch}{1.0}\begin{array}{c}1 \\1\end{array}\right) = \tfrac{1}{\sqrt{2}}\left(\renewcommand{\arraystretch}{1.0}\begin{array}{c}1 \\\I\end{array}\right) = |+\rangle\\
   \langle R | Q_R & = \tfrac{1}{\sqrt{2}}(1,-1) \left[\renewcommand{\arraystretch}{1.0}\begin{array}{cc}1 & 0 \\0 & \I\end{array}\right] = \tfrac{1}{\sqrt{2}}(1,-\I) = \langle+| \;,
\end{align}
meaning that in this direction we excite and interrogate with $\hat e_+$ microwaves (i.e., launch from the $|+\rangle$ state and project onto the $\langle+|$ state.)  By reversing direction, but retaining the same set of polarizers and quarter-wave tubes, we switch to
\begin{align}
    Q_R |R\rangle & = \left[\renewcommand{\arraystretch}{1.0}\begin{array}{cc}1 & 0 \\0 & \I\end{array}\right] \cdot \tfrac{1}{\sqrt{2}}\left(\renewcommand{\arraystretch}{1.0}\begin{array}{c}1 \\-1\end{array}\right) = \tfrac{1}{\sqrt{2}}\left(\renewcommand{\arraystretch}{1.0}\begin{array}{c}1 \\-\I\end{array}\right) = |-\rangle\\
   \langle L | Q_L & = \tfrac{1}{\sqrt{2}}(1,1) \left[\renewcommand{\arraystretch}{1.0}\begin{array}{cc}1 & 0 \\0 & \I\end{array}\right] = \tfrac{1}{\sqrt{2}}(1,\I) = \langle-| \;,
\end{align}
and therefore excite and interrogate with $\hat e_-$ microwaves.  As mentioned above, if we remove the resonator, so that the two quarter-wave tubes are connected in series,  the linearly polarized $|L\rangle$ and $|R\rangle$ states are rotated 90$^\circ$ and project perfectly onto the $\langle R|$ and $\langle L |$ states:
\begin{align}
    \langle R | Q_R  Q_L |L\rangle & = \langle + | + \rangle = 1\\
    \langle L | Q_L  Q_R |R\rangle & = \langle - | - \rangle = 1\;.   
\end{align}

\begin{figure}[t]
\includegraphics[width = 0.75 \columnwidth]{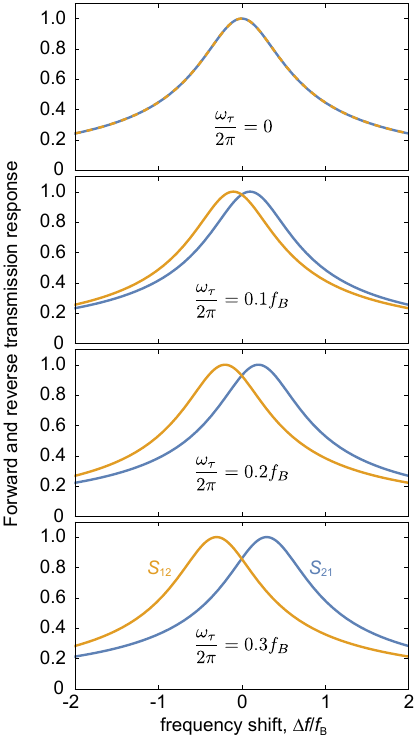}
\caption{Forward and reverse transmission amplitudes, $S_{21}$ and $S_{12}$, in the strong TRSB limit, as a function of measurement frequency, $f$, in units of the resonator bandwidth, $f_B$. The degree of TRSB is parameterized by $\omega_\tau/2\pi$, again expressed in units of $f_B$. TRSB is the only perturbation present in the resonator, and leads to visible frequency splitting and breaking of reciprocity.}
\label{fig:strong_TRSB}
\end{figure}

\subsection{Strong TRSB}

We first consider the case in which TRSB is the only perturbation present in the resonator.  This is conceptually the simplest situation, and clearly illustrates how TRSB appears as a breaking of reciprocity when the resonator is probed by circularly (or elliptically) polarized microwaves.  This limit corresponds, for example, to measuring a magnetic insulator, such as yttrium iron garnet (YIG), in which the degree of TRSB (i.e, the sample magnetization) can be tuned by an externally applied field.  The perturbation vector is $\vec \Omega = (0,\omega_\tau,0)$, implying that the angles on the Bloch sphere are $\theta = \phi = \pi/2$, and that $|\Omega| = \omega_\tau$. Using results from the previous section, the forward transmission amplitudes are
\begin{equation}
    \begin{split}
        A_\pm & = \langle R | Q_R |\psi_\pm \rangle \langle \psi_\pm | Q_L | L\rangle\\
            & = \tfrac{1}{2}\left(1 \pm \omega_\tau/|\Omega|\right)\\
            & = \tfrac{1}{2}\left(1 \pm 1\right)\\
            & = 1,0.
    \end{split}
\end{equation}
The reverse transmission amplitudes are
\begin{equation}
    \begin{split}
        \overline A_\pm & = \langle L | Q_L |\psi_\pm \rangle \langle \psi_\pm | Q_R | R\rangle\\
            & = \tfrac{1}{2}\left(1 \mp \omega_\tau/|\Omega|\right)\\
            & = \tfrac{1}{2}\left(1 \mp 1\right)\\
            & = 0, 1.
    \end{split}
\end{equation}
That is, in the forward direction, transmission occurs purely through the $|\psi_+\rangle$ state, and in the reverse direction, purely through the $|\psi_-\rangle$ state.  The breaking of reciprocity is immediately apparent, and the magnitude of TRSB is quantified by the frequency splitting between the $|\psi_+\rangle$ and $|\psi_-\rangle$ modes.  The forward and reverse transmission amplitudes are plotted in Fig.~\ref{fig:strong_TRSB}, for varying degrees of TRSB. The effects are very apparent when $\omega_\tau$ is of the order of the resonant bandwidth, but raise an important question that we address in the next section: how best to detect TRSB when $\omega_\tau$ is small compared to the bandwidth.

\subsection{Weak TRSB}

We now consider what happens when the $\omega_\tau$ perturbation is small compared to the resonant bandwidth.  We would generally expect to be in this limit when probing spontaneous TRSB, e.g., in unconventional superconductors such as UPt$_3$ and \sro.  In that case, it is beneficial to introduce a quadrupolar distortion similar in magnitude to the cavity bandwidth, as it alleviates the problem of how to resolve two closely overlapping resonances, by deliberately splitting them apart.  In practice, this can be achieved either by adjusting the sample position with respect to the cylinder axis of the resonator, or by slightly squashing the resonator out of round.

\begin{figure}[t]
\includegraphics[width = 0.75 \columnwidth]{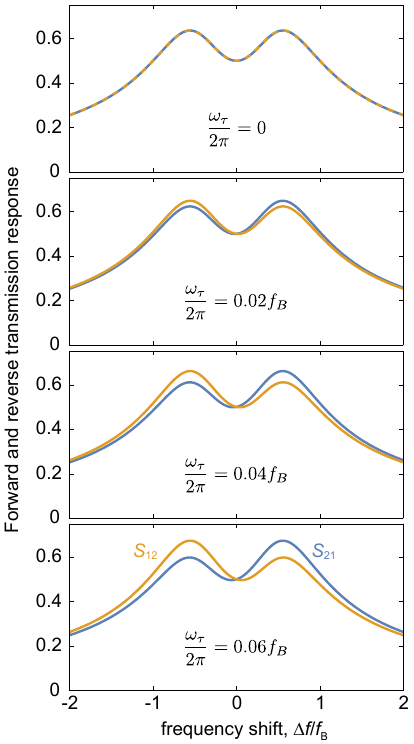}
\caption{Forward and reverse transmission amplitudes, $S_{21}$ and $S_{12}$, in the weak TRSB limit, as a function of measurement frequency, $f$, in units of the resonator bandwidth, $f_B$. The degree of TRSB is parameterized by $\omega_\tau/2\pi$, again expressed in units of $f_B$. In this case, the resonator modes are deliberately detuned by quadrupolar (shape distortion) perturbations of magnitude $\pm f_B/2$.  This makes weak TRSB visible via a first-order breaking of reciprocity, even when the change in frequency splitting is small compared to the resonator bandwidth and second order in $\omega_\tau/|\Omega|$.}
\label{fig:weak_TRSB}
\end{figure}

Again using results from above, we have forward transmission amplitudes
\begin{equation}
    \begin{split}
        A_\pm & = \langle R | Q_R |\psi_\pm \rangle \langle \psi_\pm | Q_L | L\rangle\\
            & = \tfrac{1}{2}\left(1 \pm \omega_\tau/|\Omega|\right)\;,
    \end{split}
\end{equation}
and reverse transmission amplitudes 
\begin{equation}
    \begin{split}
        \overline A_\pm & = \langle L | Q_L |\psi_\pm \rangle \langle \psi_\pm | Q_R | R\rangle\\
            & = \tfrac{1}{2}\left(1 \mp \omega_\tau/|\Omega|\right)\;.
    \end{split}
\end{equation}
The onset of weak TRSB is illustrated in Fig.~\ref{fig:weak_TRSB}. Breaking of reciprocity is quite apparent, and is well characterized using the ratiometric measure given in Eq.~\ref{eq:ratiometric_measure}, where it appears as a first-order change proportional to $\omega_\tau/|\Omega|$.

\section{Sensitivity estimates}
\label{sec:sensistivity}

To estimate the sensitivity of the TRSB resonator to changes in Kerr angle, we assume the limit of weak TRSB, where $\omega_\tau \ll |\Omega|$, and introduce a quadrupolar distortion of order half the resonator bandwidth, so that $|\Omega|/2\pi = f_B/2$.  It is this connection to the resonant bandwidth that will ultimately define the resolution of the measurement, and allow sensitivity to be boosted by increasing resonator quality factor, $Q$.

We will focus on the ratiometric measure of $\omega_\tau$ defined in Eq.~\ref{eq:ratiometric_measure} and will assume that the four amplitudes, $(A_\pm, \overline A_\pm)$, can be measured with similar fractional resolution, $\delta A/A$.  In that case
\begin{align}
    \delta\left(\frac{A_+}{A_-}\times \frac{\overline A_-}{\overline A_+}\right) & = \delta\left(\frac{\left(1 + \omega_\tau/|\Omega|\right)^2}{\left(1 - \omega_\tau/|\Omega|\right)^2}\right)\\
    \Rightarrow 4 \frac{\delta A}{A} & = 4 \frac{\delta \omega_\tau}{|\Omega|}\;.
\end{align}
That is,
\begin{equation}
    \delta \omega_\tau = |\Omega| \frac{\delta A}{A} = \pi f_B \frac{\delta A}{A}\;.
\end{equation}
Relating this  to Kerr angle, we have
\begin{align}
    \delta \theta_K & = \frac{2}{\Gamma Z_0} \delta \omega_\tau\\
        & = \frac{4 V_\mathrm{eff}}{A_\mathrm{sample}} \frac{1}{c} \delta \omega_\tau \\
        & = \frac{4 V_\mathrm{eff}}{A_\mathrm{sample}} \frac{1}{c} \pi f_B \frac{\delta A}{A} \\
        & = \frac{4 V_\mathrm{eff}}{A_\mathrm{sample}} \frac{f0}{c} \frac{\pi}{Q} \frac{\delta A}{A} \\
        & = \frac{4 V_\mathrm{eff}}{A_\mathrm{sample} \lambda} \frac{\pi}{Q} \frac{\delta A}{A}\;.
\end{align}
We can now evaluate the sensitivity in various limits.  As expected, the minimum detectable change in Kerr angle improves with increasing resonator $Q$.  From previous experience operating microwave resonators at dilution fridge temperatures \cite{Truncik:2013hr}, a typical fractional resolution for amplitude is $\delta A/A = 10^{-3}$, for a single-sweep measurement made at low microwave power.  If a normal-metal resonator is required (e.g., to allow external magnetic field to be applied, without flux trapping) then a typical quality factor would be $Q = 3 \times 10^4$, for a copper resonator.  This could potentially be improved to $Q = 10^6$ for a copper-shielded dielectric resonator \cite{Huttema:2006p344} while still allowing external field to be applied.  Alternatively, $Q = 10^6$ is easily attainable using superconducting resonators. The final variable is the size of the sample.  As outlined in Sec.~\ref{sec:circ_pol_microwave_modes}, if the sample is large enough to form one end wall of the resonator, the factor $4 V_\mathrm{eff}/A_\mathrm{sample} \lambda \approx 1$, on quite general grounds.  In this limit
\begin{align}
    \delta \theta_K & = \frac{\pi}{Q} \frac{\delta A}{A}\;.
\end{align}
This works out to a single-sweep Kerr-angle resolution of 100~nanoradians for a normal metal resonator \mbox{($Q = 3 \times 10^4$)} and a resolution of 3~nanoradians for a superconducting or dielectric resonator \mbox{($Q = 10^6$)}.

In the more common situation where the sample is a \mbox{mm-sized} single crystal, the resolution will be lower by approximately the ratio of sample area to end-wall area.  Assuming a TE$_{111}$ resonator operating around 16~GHz, with a diameter of 1.25~cm and a height of 1.8~cm, and a two-sided sample area $A_\mathrm{sample} = 2$~mm$^2$, we obtain a single-sweep sensitivity of $\delta \theta_K \approx 10$~microradians for $Q = 3 \times 10^4$, and $\delta \theta_K \approx 300$~nanoradians for $Q = 10^6$.  We emphasize that these are single-sweep sensitivities, and can be further improved using signal-averaging techniques.

\section{Conclusions}

Motivated by a desire to probe spontaneous TRSB at photon energies comparable to the superconducting gaps of unconventional superconductors, such as UPt$_3$ and \sro, we have explored the microwave-frequency electrodynamics of materials that break time-reversal symmetry, showing that TRSB generically leads to a difference in surface reactance for left- and right-circularly polarized microwave fields.  This can be probed using cavity perturbation techniques, as long as a doubly degenerate resonator mode is employed, such as the TE$_{111}$ mode of a cylindrical cavity.  While TRSB immediately lifts the degeneracy of the modes and imparts a characteristic frequency splitting, parameterized by $\omega_\tau$, it is in general difficult to distinguish splitting due to TRSB from spurious frequency splitting arising, for example, from shape distortions that break the cylindrical symmetry of the resonator.  A more detailed treatment shows that when the resonator is interrogated using left- and right-circularly polarized microwaves, TRSB is uniquely associated with breaking of reciprocity and appears as a difference in the forward and reverse transmission responses of the resonator.  This provides a crucial test that distinguishes true polar Kerr effects from trivial effects such as linear birefringence.  It also provides a measurement method that continues to work even when the effects of TRSB are much smaller than those of shape distortion and cavity dissipation (i.e., when $\omega_\tau \ll |\Omega|, f_B$). 

The microwave-resonator-based technique should provide insights complementary to work carried out in the near infrared using the zero-loop-area Sagnac interferometer developed by the Kapitulnik group \cite{Xia:2006p36}.  In both cases, the techniques are sensitive to the nonreciprocal phase shift $\phi_\mathrm{nr} = 2 \theta_K$ that arises between left and right circularly polarized electromagnetic waves reflected from the surface of a sample that breaks time-reversal symmetry.  In the microwave case, a high quality factor enhances the interaction between the sample and the circularly polarized microwaves.  This leads, in the limit where sample is larger than the spot size (i.e., resonator diameter), to sensitivities comparable to the zero-loop-area Sagnac.  

\begin{acknowledgments}
We acknowledge A.~Kapitulnik for suggesting that TRSB be measured with microwaves.  We acknowledge initial attempts by K.~J.~Morse to realize that goal.  We are grateful to J.~S.~Dodge, P.~J.~Hirschfeld, V.~Mishra and J.~E.~Sonier for useful discussions.  Financial support for this work was provided by the Natural Science and Engineering Research Council of Canada. 
\end{acknowledgments}

\appendix*

\section{TE$_\mathbf{111}$ mode of a cylindrical resonator}
\label{sec:appendix}

The magnetic field $\vec H = (H_r, H_\phi, H_z)$ of the TE$_{111}$ mode of a cylindrical cavity resonator can be written in cylindrical coordinates as
\begin{align}
    \begin{split}
        H_r & = H_0\,\E^{\I \phi}\,J_1^\prime(k_r r) \cos(k_z z)\\
        H_\phi & = H_0\,\I \E^{\I \phi}\,\frac{1}{k_r r}J_1(k_r r) \cos(k_z z)\\
        H_z & = H_0\,\E^{\I \phi}\,\frac{k_r}{k_z}J_1(k_r r) \sin(k_z z)\;,
    \end{split}
    \label{eq:TE111_fields}
\end{align}
where $J_1(x)$ is a Bessel function of the first kind, \mbox{$k_r = 1.8412/a$} and $k_z = \pi/d$ are chosen to satisfy the boundary conditions at $r = a$ and $z = d$, and $a$ and $d$ are the radius and height of the cavity, respectively.  Taking the real and imaginary parts of the fields generates the two orthogonal linear polarizations of the TE$_{111}$ mode in the $z$ basis.  

The resonator constant $\Gamma$ is related to the effective cavity volume
\begin{align}
    \begin{split}    
        V_\mathrm{eff} &= \frac{1}{H_\mathrm{sample}^2} \int_V H^2 dV \\
                     &= \frac{1}{\big|H_\mathrm{sample}\big|^2} \int_V \big|H\big|^2 dV\;,
    \end{split}
    \label{eq:effective_volume_appendix}
\end{align}
where the second form holds if the complex form of the resonator fields in Eq.~\ref{eq:TE111_fields} is used directly.  If the sample is placed at the centre of the end wall of the cavity (i.e., at $r = z = 0$) then the magnetic field at the sample in each of the polarizations is $H_\mathrm{sample} = H_0/2$, which is equivalent in the complex version of Eq.~\ref{eq:effective_volume_appendix} to $\left|H_\mathrm{sample}\right| = H_0/\!\sqrt{2}$.
Working with the fields in complex form, the azimuthal dependence immediately cancels out when forming $|H|^2$, leading to
\begin{align}
    \begin{split}
        V_\mathrm{eff} &= 4 \pi \!\!\int_0^d \!\!\!\!\!dz \!\!\int_0^a \!\!\!\!\! r dr \!\left[\cos^2(k_z z)\left(J_1^{\prime 2}(k_r r) + \frac{1}{(k_r r)^2} J_1^2(k_r r) \right) \right.\\
                    & + \left.\sin^2(k_z z) \frac{k_r}{k_z} J_1^2(k_r r)  \right]\;.
    \end{split} 
\end{align}
Each term of the $z$ integral contributes a factor of $d/2$.  To carry out the $r$ integral we make the substitution \mbox{$x = k_r r = 1.8412 r/a$}:
 \begin{align}
    \begin{split}
        V_\mathrm{eff} & = \frac{2 \pi d}{k_r^2} \!\!\int_0^{k_r a}\left[ x J^{\prime 2}_1(x) + \left(\frac{1}{x} + \left(\frac{k_r}{k_z}\right)^2 \!\!x\right) J_1^2(x) \right] dx\\
         & = \frac{2 \pi a^2 d}{(1.8412)^2} \times 0.405 \left[1 + \left(\frac{k_r}{k_z}\right)^2\right]\\
         & = 0.239\,V_\mathrm{cavity} \left[1 + \left(\frac{1.8412 d}{\pi a}\right)^2\right] \\
         & = 0.239\,V_\mathrm{cavity} \left(1 + 0.343\frac{d^2}{a^2}\right)\;.
    \end{split} 
\end{align}
A typical value of the aspect ratio for a TE$_{111}$ cavity is \mbox{$d/a = 3$}, as this keeps the the TE$_{111}$ and TM$_{010}$ modes well separated.  Then $V_\mathrm{eff} = 0.98 V_\mathrm{cavity}$, where \mbox{$V_\mathrm{cavity} = \pi a^2 d$}.


%

\end{document}